\documentclass[preprint,aps]{revtex4}

\usepackage{graphicx}

\begin{document}

\title{Tunable Dirac Fermion Dynamics in Topological Insulators}

\author{Chaoyu Chen$^{1}$, Zhuojin Xie$^{1}$, Ya Feng$^{1}$, Hemian Yi$^{1}$, Aiji Liang$^{1}$, Shaolong He$^{1}$, Daixiang Mou$^{1}$, Junfeng He$^{1}$, Yingying Peng$^{1}$, Xu Liu$^{1}$, Yan Liu$^{1}$, Lin Zhao$^{1}$, Guodong Liu$^{1}$, Xiaoli Dong$^{1}$, Jun Zhang$^{1}$, Li Yu$^{1}$, Xiaoyang Wang$^{2}$, Qinjun Peng$^{2}$, Zhimin Wang$^{2}$, Shenjin Zhang$^{2}$, Feng Yang$^{2}$, Chuangtian Chen$^{2}$, Zuyan Xu$^{2}$ and X. J. Zhou$^{1,*}$}

\affiliation{
\\$^{1}$Beijing National Laboratory for Condensed Matter Physics, Institute of Physics, Chinese Academy of Sciences, Beijing 100190, China
\\$^{2}$Technical Institute of Physics and Chemistry, Chinese Academy of Sciences, Beijing 100190, China.
\\$^{*}$Corresponding author: XJZhou@aphy.iphy.ac.cn.
}

\date{February 1, 2013}
%
%

\begin{abstract}

Three-dimensional topological insulators are characterized by insulating bulk state and metallic surface state involving Dirac fermions that behave as massless relativistic particles. These Dirac fermions are responsible for achieving a number of novel and exotic quantum phenomena in the topological insulators and for their potential applications in spintronics and quantum computations. It is thus essential to understand the electron dynamics of the Dirac fermions, i.e., how they interact with other electrons, phonons and disorders. Here we report super-high resolution angle-resolved photoemission studies on the Dirac fermion dynamics in the prototypical Bi$_{2}$(Te,Se)$_{3}$ topological insulators. We have directly revealed signatures of the electron-phonon coupling in these topological insulators and found that the electron-disorder interaction is the dominant factor in the scattering process. The Dirac fermion dynamics in  Bi$_2$(Te$_{3-x}$Se$_x$) topological insulators can be tuned by varying the composition, x, or by controlling the charge carriers.  Our findings provide crucial information in understanding the electron dynamics of the Dirac fermions in topological insulators and in engineering their surface state for fundamental studies and potential applications.
\end{abstract}


\maketitle

Topological insulators represent a new state of matter that has an insulating bulk state but a conducting surface$\slash$edge state which hosts an unique spin texture and a linear dispersion relation compatible with massless Dirac fermions\cite{RevMoore1,RevZSC,RevHasan}.  In addition to their potential applications in spintronics and quantum computing\cite{RevQXL, RevMoore2}, the topological insulators also provide a promising platform for realizing a number of novel and exotic quantum phenomena including quantum spin Hall effect\cite{SCZhangQSHEHgTe}, fractional charge and quantized current\cite{QXLFracCharge}, dynamical axion field\cite{LRDAxion}, magnetic monopole\cite{QXLMonopole}, Majorana fermions\cite{FLMajorana, FLTSCCuBiSe} and quantum anomalous Hall effect\cite{YRQAHE}. Some of the phenomena have been observed experimentally in two-dimensional topological insulators, such as quantum spin Hall effect\cite{MolenkampQSHEHgTe} and quantum Hall effect\cite{MolenkampQHEHgTe} in the HgTe quantum wells\cite{SCZhangQSHEHgTe}. However, in the case of three-dimensional topological insulators like Bi$_{2}$(Te, Se)$_{3}$ family\cite{FL3DTIs,ZhongFangBiSe, HasanBiSeARPES,YLChen}, in spite of intensive efforts from transport experiments\cite{OngQuanOsci, ButchBiSeTransport, AndoDiracHole, FisherQuanLim, Ando2DQO}, attempts to observe electronic transport signatures of the surface Dirac fermions have been hampered by  a couple of obstacles. First, contrary to usual expectation that the bulk is insulating in topological insulators, the presence of defects in Bi$_{2}$(Te, Se)$_{3}$ system makes the bulk conducting\cite{CavaPtypeBiSe} and the transport properties are then dominated by the bulk state instead of the desired surface state\cite{ButchBiSeTransport}. Second, in several cases where the surface transport signatures are identified, more than one frequency or one type of charge carriers contribute to the quantum oscillation\cite{AndoDiracHole, FisherQuanLim, Ando2DQO}, indicating the complexity of sample surface exposed to atmosphere or protection layers\cite{XJZhouBiTeSeARPES}. Third, the transport mobility of charge carriers in the Bi$_{2}$(Te, Se)$_{3}$ topological insulators, $\mu^{'} \sim 10^{4}\ cm^{2}V^{-1}s^{-1}$\cite{OngQuanOsci, Ando2DQO, FuhrerEPCTransport}, is not sufficiently high compared to those from HgTe quantum wells($1.5 \times 10^{5}\ cm^{2}V^{-1}s^{-1}$) \cite{MolenkampQSHEHgTe, MolenkampQHEHgTe} and graphene($2.3 \times 10^{5}\ cm^{2}V^{-1}s^{-1}$) \cite{NovoselovGraphene, KimGraphene, StormerGraphene}. Although strict electron backscattering is forbidden in topological insulators due to time reversal symmetry invariance,  it is important to uncover other scattering channels that limit the transport mobility of the two-dimensional Dirac fermions\cite{FuhrerEPCTransport}.

The charge transport properties are dictated by the electron dynamics in materials. In particular, the scattering of electrons by other entities such as other electrons, phonons and disorder,  determines the lifetime of the quasiparticles, and thus the electron mobility and mean free path in the transport process\cite{Grimvall}. Angle-resolved photoemission spectroscopy (ARPES) has become a powerful tool to directly probe such many-body effects\cite{RevDamascelli,RevXJZhou}.  Under the sudden approximation, ARPES measures a single particle spectral function

\begin{equation}\label{}
{A(k,\omega) = \frac{1}{\pi} \frac{\mid Im\Sigma(k,\omega)\mid}{[\omega-\epsilon_{k}-Re\Sigma(k,\omega)]^{2} + [Im\Sigma(k,\omega)]^{2}}}
\end{equation}

\noindent with {\it k} and $\omega$ representing electron momentum and energy, respectively. The real and imaginary parts of electron self-energy, $Re\Sigma(k,\omega)$ and  $Im\Sigma(k,\omega)$, carry all the information about the band renormalization and quasiparticle lifetime in an interacting system\cite{RevDamascelli, RevXJZhou}.

There have been a few investigations on the dynamics of Dirac fermions in Bi$_{2}$(Te, Se)$_{3}$ topological insulators using ARPES\cite{HofmannEPCEDS, VallaWeakEPC} or helium atom surface scattering probes\cite{BatanounyHASSEPC}. However, the results obtained so far are not consistent. While the electron-phonon coupling strength ($\lambda$) was reported to be 0.25 in Bi$_2$Se$_3$ from temperature-dependent photoemission measurements\cite{HofmannEPCEDS}, it was claimed that the electron-phonon coupling is exceptionally weak in Bi$_2$Se$_3$ in another measurement\cite{VallaWeakEPC}. The inconsistency comes presumably from two issues related to the ARPES study on many-body effects in topological insulators. First, the coexistence of bulk  bands and surface bands near the Fermi level, due to the bulk doping by defects, complicates the identification of weak band renormalization effects in the surface bands. Second, the small momentum space and energy window of the Dirac fermions and the weak many-body effects require high precision ARPES measurements.  The momentum space occupied by the surface state, $\sim\pm$ 0.1 ${\AA}^{-1}$ for Bi$_{2}$Se$_{3}$ (Fig. 1A) and $\sim\pm$ 0.04 ${\AA}^{-1}$ for p-type Bi$_{2}$Te$_{3}$ (Fig. 2A), is rather small \cite{HasanBiSeARPES, ZXShenBiTeARPES, XJZhouBiTeSeARPES}. To overcome these problems, it is ideal to first suppress the interference of the bulk bands in the measurements.  Since the as-grown Bi$_{2}$Se$_{3}$ is electron-doped (n-type) while Bi$_{2}$Te$_{3}$ can be hole-doped (p-type)\cite{XJZhouBiTeSeARPES}, balancing the proportion of selenium and tellurium in Bi$_{2}$(Se,Te)$_{3}$ system may neutralize the bulk carrier doping and reduce the effect of the bulk bands on the surface state.  Furthermore, vacuum ultra-violet (VUV) laser-based ARPES\cite{XJZhouLaserARPES} can help suppress the bulk bands\cite{XJZhouBiTeSeARPES}, in addition to its super-high energy and momentum resolutions (see Laser-ARPES measurements and Fig. S2 in Supplementary Materials) that are desirable in studying many-body effects in topological insulators.

The Bi$_2$Se$_3$ sample shows sharp surface state band (Fig. 1A) that facilitates the investigation of the Dirac fermion dynamics in the material. The momentum distribution curves (MDCs) at different binding energies (Fig. 1B) show well-defined sharp peaks that can be fitted by Lorentzian lineshape to obtain peak position and peak width (Full-Width-at-Half-Maximum, $\Gamma_{MDC}$) (see Fig. S3 in Supplementary Materials).  The obtained MDC peak position as a function of the binding energy gives the measured dispersion relation (left panels of Fig. 1C and 1D). We assume a bare band as a linear line that connects the two points on the measured dispersion, one at the Fermi level and the other at 100 meV binding energy. This is reasonable within such a small energy window and particularly for the Dirac fermions that shows a linear dispersion over a large energy range. The effective real part of the electron self-energy (Re$\Sigma$) is obtained by subtracting the measured dispersion with the bare band, as shown in right panels of Fig. 1C and Fig. 1D (black empty circles). The imaginary part of the electron self-energy (Im$\Sigma$) is obtained from the fitted MDC width: Im$\Sigma$=$\Gamma_{MDC}$$\times$$v_0$$\slash$2 with  $v_0$ being the Fermi velocity of the bare band (right panels of Fig. 1C and Fig. 1D, blue empty squares)\cite{RevXJZhou}.

The extraction of the electron self-energy provides full information for understanding the electron dynamics in a  material where the electron scattering usually involves three main processes: electron-electron interaction, electron-phonon coupling, and electron-disorder interaction\cite{VallaMo,JensenBe}. The electron-disorder interaction is characterized by the imaginary part of the electron self-energy at zero energy (Fermi level) since the contribution from the electron-electron interaction and electron-phonon coupling approaches zero at the zero energy at very low temperature. The electron-phonon interaction is characterized by the band renormalization (usually a kink in dispersion, corresponding to a peak in the real part of the electron self-energy) and a drop in the imaginary part of the electron self-energy. As exemplified in Fig. 1 for Bi$_2$Se$_3$, there are clear indications of electron-phonon coupling revealed, such as the kink in dispersions (left panels of Fig. 1C and Fig. 1D), peak in the real part of the electron self-energy and a drop in the imaginary part of the electron self-energy (right panels of Fig. 1C and Fig. 1D). The characteristic energy scale of the phonons involved ($\omega_0$) is determined by the peak position of the real part of the electron self-energy; in the case of Bi$_2$Se$_3$, it is $\sim$18 meV as marked by arrows in Fig. 1C and 1D. The electron-phonon coupling strength, $\lambda$, can be determined by the slope of the line near the zero energy in the real part of the electron self-energy. As shown by red dashed line in the right panel of Fig. 1C, it is determined to be $\lambda$$\sim$0.17 for Bi$_2$Se$_3$. The electron-electron interaction manifests itself more clearly in the imaginary part of the electron self-energy. In a typical Fermi liquid case, its contribution is proportional to the (E-E$_F$)$^2$: Im$_{ee}$=$\beta$(E-E$_F$)$^2$. Since the electron-phonon contribution is constant above the characteristic phonon scale, the deviation from the constant above the phonon scale gives a measure of the electron-electron interaction involved. For Bi$_2$Se$_3$ and the Bi$_2$(Se,Te)$_3$ series, we find that the electron-electron contribution is weak ($\beta$$\sim$0.014 for Bi$_2$Se$_3$), particularly when compared with the dominant electron-disorder interaction and the electron-phonon coupling. Therefore, in the following, we will mainly focus on the electron-phonon coupling and electron-disorder interaction.

As shown in Fig. 1, our measurements clearly indicate that there is an obvious electron-phonon coupling ($\lambda$$\sim$0.17) in the electron dynamics of the Dirac fermions in Bi$_2$Se$_3$. It is nearly isotropic because similar behaviors are observed along the $\overline{\Gamma}$--$\overline{K}$ (Fig. 1C) and $\overline{\Gamma}$--$\overline{M}$ (Fig. 1D) momentum cuts.  The characteristic energy scale, $\sim$18 meV, is consistent with the highest-frequency phonon mode ($\mathbf{\emph{A}^{2}_{1g}}$) at $\sim$ 20 meV in Bi$_2$Se$_3$ from both theoretical calculations\cite{WCPhononSpectrum} and Raman measurements\cite{BeckerRaman} (see Fig. S7 in Supplementary Materials). This is in  strong contrast to a recent report where exceptionally weak electron-phonon coupling was claimed and no dispersion kink was resolved in Bi$_{2}$Se$_{3}$\cite{VallaWeakEPC}. We believe that the clear manifestation of electron-phonon coupling in our measurements is mainly due to much improved instrumental resolution we have achieved.

The electron-phonon coupling is extremely weak in the p-type Bi$_2$Te$_3$ sample, as seen in Fig. 2. This sample was prepared under similar condition as the above Bi$_2$Se$_3$ sample and it is p-type because the Fermi level intersects with the bulk valence band\cite{XJZhouBiTeSeARPES}. Even measured with such a super-high instrumental resolution (see Fig. S2 in Supplementary Materials), we do not resolve any clear signature of electron-phonon coupling. The measured dispersion is basically a straight line, corresponding to a real part of electron self-energy that is nearly zero (black empty circles in right panels of Fig. 2C and Fig. 2D). There is no clear drop in the imaginary part of the electron self-energy (blue empty squares in right panels of Fig. 2C and Fig. 2D). All these observations point to an extremely weak electron-phonon coupling in the p-type Bi$_2$Te$_3$ sample. In the mean time, the p-type Bi$_2$Te$_3$ also exhibits a rather weak electron-disorder scattering when compared with Bi$_2$Se$_3$. As seen from the right panel of Fig. 2C, the imaginary part of the electron self-energy at zero energy is only 5.6 meV  along the $\overline{\Gamma}$--$\overline{K}$ direction. This is significantly lower than 37 meV along the same direction in Bi$_2$Se$_3$ topological insulator (right panel of Fig. 1C).

The dramatically different behaviors between Bi$_2$Se$_3$ (Fig. 1) and p-type Bi$_2$Te$_3$ (Fig. 2) motivated us to further investigate the composition dependence of the Dirac fermion dynamics in the Bi$_2$Te$_{3-x}$Se$_x$ series. We observe a systematic evolution of band structure, electron-phonon coupling, and electron-disorder scattering, with the change of composition x in Bi$_2$Te$_{3-x}$Se$_x$  (Fig. 3). Note that all these samples are prepared under similar conditions. First, the location of the Dirac point (E$_{D}$) moves monotonically to higher binding energy with increasing  $x$. For 0 $\leq x \leq$ 0.05, the samples are p-type because the Fermi level intersects with the bulk valence band.  For 0.15 $\leq x \leq$ 1.5, the Fermi level intersects with the surface state bands only. For 1.5 $< x \leq$ 3, the samples become n-type because the Fermi level intersects with the bulk conduction band. Note that the Dirac point exhibits a big jump when x changes from 0.05 to 0.15.  Second, there is an overall increase of electron-disorder interaction with increasing x in Bi$_2$Te$_{3-x}$Se$_x$. This can be seen from the increase of the imaginary part of the electron self-energy and hence its value at zero energy with increasing x. This indicates that the Bi$_2$Te$_{3-x}$Se$_x$ system gets more disordered with increasing x. Third, there is also an overall increase in the electron-phonon coupling strength with increasing x in Bi$_2$Te$_{3-x}$Se$_x$. Although no indication of electron-phonon coupling is observed in p-type Bi$_2$Te$_3$, a slight substitution of Te by Se, with x=0.05, already renders a clear observation of electron-phonon coupling in Bi$_2$Te$_{2.95}$Se$_{0.05}$, as seen in the appearance of a peak in the real part of the electron self-energy (black circles in Fig. 3C2). The electron-phonon coupling strength increases from nearly zero for p-type Bi$_2$Te$_3$ to $\lambda$$\sim$0.17 for Bi$_2$Se$_3$.

It is interesting that for the same Bi$_2$Te$_3$ nominal composition, when prepared under different conditions, samples can become either p-type or n-type that exhibit dramatic difference in their Dirac fermion dynamics. As shown in Fig. 3B,  the Dirac point of the n-type Bi$_2$Te$_3$ lies nearly 310 meV below the Fermi level which differs significantly from the 85 meV location in the p-type Bi$_2$Te$_3$ (Fig. 3A1). Particularly, the n-type Bi$_2$Te$_3$ exhibits a pronounced electron-phonon coupling, as seen from the obvious peak in the real part of the electron self-energy (black circles in Fig. 3D) and a drop in the imaginary part of the electron self-energy (blue squares in Fig. 3D). The resultant electron-phonon coupling constant is $\lambda$$\sim$0.19, which is even slightly larger than that in Bi$_2$Se$_3$ ($\lambda$$\sim$0.17). The electron-disorder scattering also increases in the n-type Bi$_2$Te$_3$ compared to the p-type one, as the value of the imaginary part of the electron self energy at zero energy becomes much larger (Fig. 3D). However, compared with Bi$_2$Se$_3$, the disorder scattering in n-type Bi$_2$Te$_3$ remains much weaker(Fig. 3C11). The first principle calculation, which expects a rather weak electron-phonon coupling in Bi$_2$Te$_3$ ($\lambda$$\sim$0.05)\cite{HuangPhononBiTe}, is hard to explain the strong electron-phonon coupling in n-type Bi$_2$Te$_3$ and particularly the disparity of electron-phonon coupling between the n-type and p-type Bi$_2$Te$_3$.  As the main difference between the n-type and p-type Bi$_2$Te$_3$ lies in the relative position between the Dirac point and the bulk bands, it is reasonable to deduce that the bulk bands may play a role in the electron dynamics of the Dirac fermions.  This is consistent with the observation that, for n-type Bi$_2$Te$_3$ (x=3), although its composition is rather different from Bi$_2$Te$_{0.15}$Se$_{2.85}$ (x=0.15), their Dirac fermion dynamics is quite similar in both the electron-phonon coupling and electron-disorder interaction (Figs. 3 and 4). The distinction of the electron-phonon coupling between the n-type and p-type Bi$_{2}$Te$_{3}$ indicates that the control of the charge carrier doping can be an alternative way to tune the Dirac fermion dynamics in topological insulators.

Fig. 4 gives a quantitative summary of the electron-phonon coupling and electron-disorder scattering in the Bi$_2$Te$_{3-x}$Se$_x$ system. The relative position between the Dirac point and the bulk band is depicted in Fig. 4A for a couple of typical compositions and both p- and n-type Bi$_2$Te$_3$. With increasing x in Bi$_2$Te$_{3-x}$Se$_x$, there is an abrupt jump in the electron-phonon coupling strength from x=0 to x=0.15, accompanied by a transition from p-type to n-type. Further increase of x leads to an overall slight increase of electron-phonon coupling (Fig. 4C). Meanwhile, the characteristic energy scale, $\omega$$_0$, also increases with x, which is consistent with the increase of the phonon frequency from Bi$_2$Te$_3$ to Bi$_2$Se$_3$ (see Fig. S7 in Supplementary Materials). This indicates that the (Te,Se) vibrations play a major role in dictating the electron-phonon coupling strength in the Bi$_2$Te$_{3-x}$Se$_x$ system. This is also consistent with the observation that the characteristic energy scale is identical in n-type and p-type Bi$_2$Te$_3$ because their phonon frequencies are quite similar (see Fig. S7 in Supplementary Materials) although they have different electron-phonon coupling strength. It is clear that electron-disorder scattering  is dominant in the scattering process. As seen from Fig. 4B, the imaginary part of the electron self-energy at zero energy Im$\Sigma$(E$_F$), which gives a good measure of the electron scattering rate, shows an overall increase from the p-type Bi$_2$Te$_3$ for x=0 to Bi$_2$Se$_3$ for x=3 in the Bi$_2$Te$_{3-x}$Se$_x$ system (red circles in Fig. 4B). The quasi-particle mean free path, $\ell^{'}$, which can be determined from the MDC width $\Gamma_{MDC}$, $\ell^{'} = 1/\Gamma_{MDC}$, first decreases abruptly from the p-type Bi$_2$Te$_3$ (x=0) to x=0.15, followed by an overall slight decrease with x in Bi$_2$Te$_{3-x}$Te$_x$ (blue circles in Fig. 3B).  The electron mobility of the Dirac fermions ($\mu^{'}$) is directly related to the quasi-particle mean free path,  $\mu^{'} = e\ell^{'}/(\hbar k_{F})$, with k$_{F}$ being the Fermi momentum. It is then estimated that the electron mobility in p-type Bi$_2$Te$_3$ is near 1000 $cm^{2}V^{-1}s^{-1}$ (see Supplementary Materials). By comparison, the electron mobility in Bi$_2$Se$_3$ and n-type Bi$_2$Te$_3$ is $\sim$70 and $\sim$100 $cm^{2}V^{-1}s^{-1}$, respectively, which are significantly lower than that of p-type Bi$_2$Te$_3$. We note that, although the electron mobility obtained from ARPES is not strictly the same as the electron mobility in the transport properties, they have good correspondence, especially for materials with circular Fermi surface and nearly isotropic electron scattering.

 The present study will provide important information for engineering the Dirac fermion dynamics in Bi$_2$Te$_{3-x}$Se$_x$ topological insulators. It clearly indicates the presence of electron-phonon coupling and dominant role of the electron-disorder scattering in dictating the Dirac fermion dynamics in the system.  The Dirac fermion dynamics can be tuned by varying either the composition or the charge carrier concentration.  Specifically for the Bi$_2$Te$_{3-x}$Se$_x$ system, our results clearly show that Bi$_2$Te$_3$ system is superior over Bi$_2$Se$_3$ in achieving higher electron mobility both due to its weak electron-phonon coupling and its weak electron-defect scattering.  To further enhance the electron mobility in searching for novel quantum phenomena, it is necessary to choose a system with weak electron-phonon coupling and reduced disorder scattering.

\noindent{\bf References and Notes}

\newpage

\begin{figure}[tbp]
\begin{center}
\includegraphics[width=1\columnwidth,angle=0]{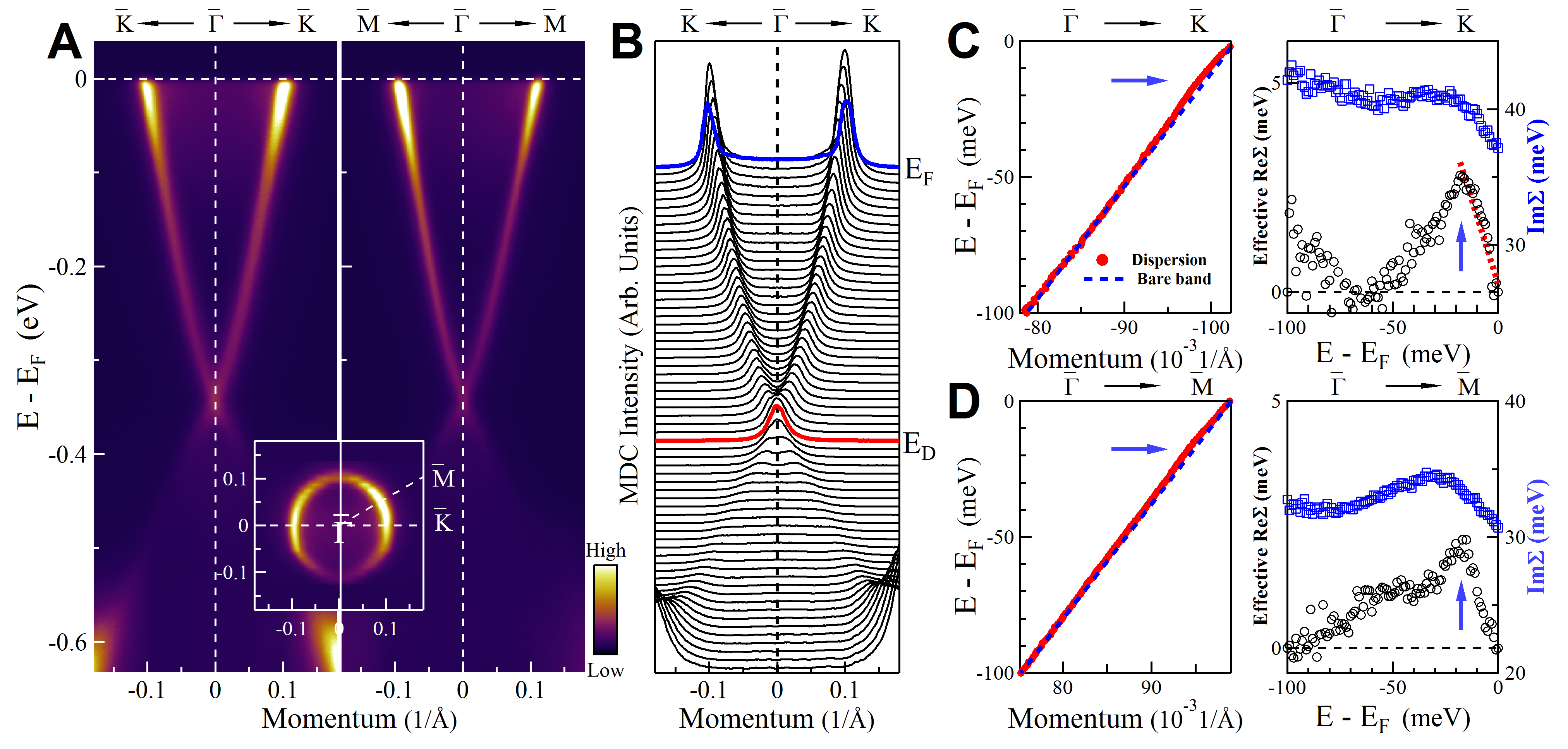}
\end{center}
\caption{Observation of electron-phonon coupling in Bi$_{2}$Se$_{3}$. (\textbf{A}) Surface state dispersions along $\overline{\Gamma}$--$\overline{K}$ (left panel) and $\overline{\Gamma}$--$\overline{M}$ (right panel) directions. The inset shows the Fermi surface of Bi$_{2}$Se$_{3}$. (\textbf{B}) Momentum distribution curves (MDCs) of the surface state dispersion along $\overline{\Gamma}$--$\overline{K}$. Red line and blue line indicate the MDCs at the Dirac point (E$_{D}$) and the Fermi level (E$_{F}$), respectively. (\textbf{C}) Dispersion and effective electron self-energy of surface state band along $\overline{\Gamma}$--$\overline{K}$ direction. The left panel shows the measured dispersion (red circles) obtained from MDC fitting.  The blue dotted line represents an empirical bare band that is a straight line connecting two points in the measured dispersion, one at E$_F$ and the other at 100 meV binding energy.  The right panel shows the effective real part of the electron self energy, $Re\Sigma(k,\omega)$ (black circles), and imaginary part of the electron self energy, $Im\Sigma(k,\omega)$ (blue squares). The red dashed line is a linear fit to the low-energy region of $Re\Sigma(k,\omega)$. Blue arrows indicate the characteristic energy scale of the electron-phonon coupling. (\textbf{D})Dispersion and effective electron self-energy of surface state band along $\overline{\Gamma}$--$\overline{M}$ direction.
}
\end{figure}

\newpage

\begin{figure}[tbp]
\begin{center}
\includegraphics[width=1\columnwidth,angle=0]{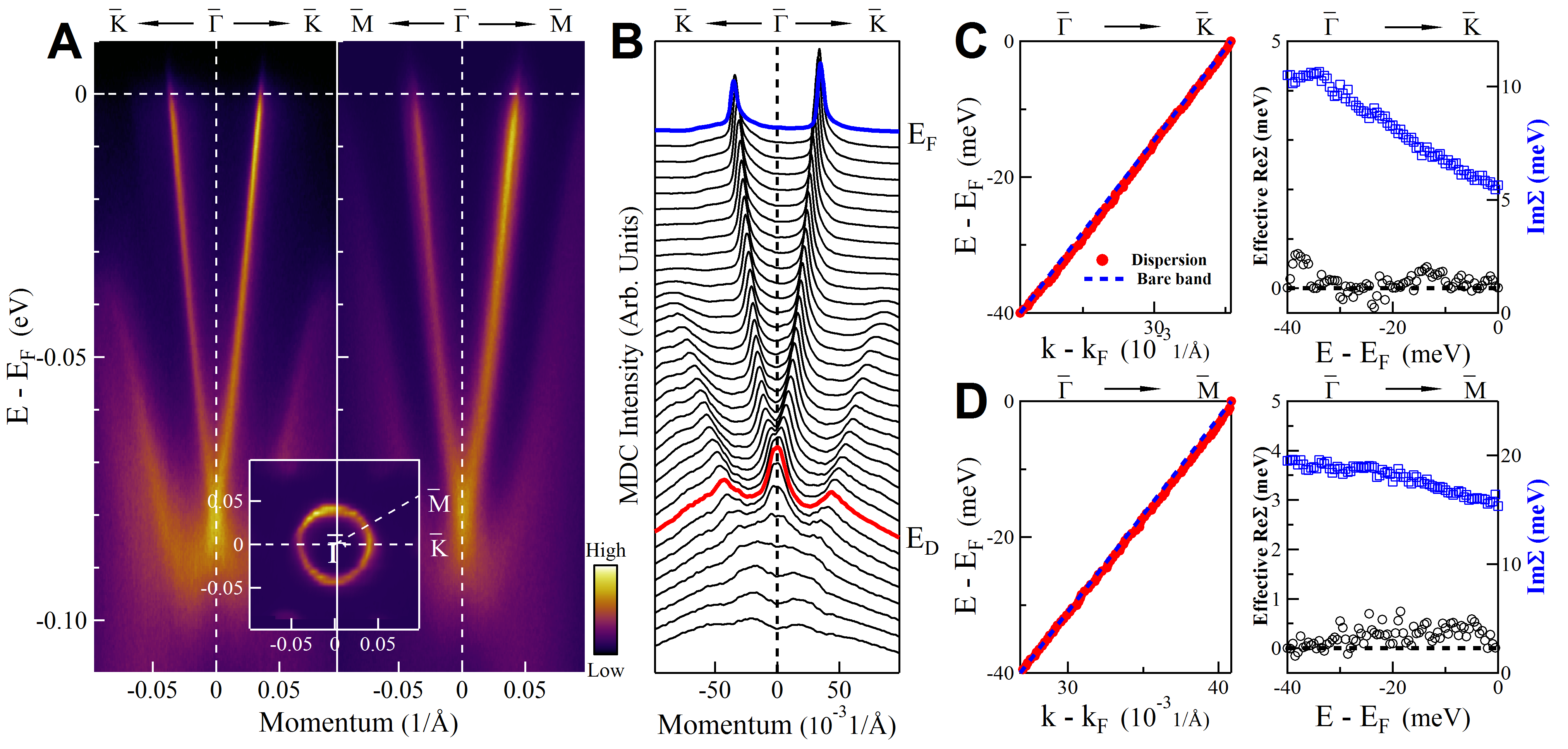}
\end{center}
\caption{Extremely weak electron-phonon coupling in p-type Bi$_{2}$Te$_{3}$.  (\textbf{A}) Surface state dispersions along $\overline{\Gamma}$--$\overline{K}$ (left panel) and $\overline{\Gamma}$--$\overline{M}$ (right panel) directions. The inset shows the Fermi surface of p-type Bi$_{2}$Te$_{3}$. (\textbf{B}) Momentum distribution curves (MDCs) of the surface state dispersion along $\overline{\Gamma}$--$\overline{K}$. Red line and blue line indicate the MDCs at the Dirac point (E$_{D}$) and the Fermi level (E$_{F}$), respectively. (\textbf{C}) Dispersion and effective electron self-energy of surface state band along $\overline{\Gamma}$--$\overline{K}$ direction. The left panel shows the measured dispersion (red circles) obtained from MDC fitting.  The blue dotted line represents an empirical bare band that is a straight line connecting two points in the measured dispersion, one at E$_F$ and the other at 40 meV binding energy.  The right panel shows the effective real part of the electron self energy, $Re\Sigma(k,\omega)$ (black circles), and imaginary part of the electron self energy, $Im\Sigma(k,\omega)$ (blue squares).  (\textbf{D})Dispersion and effective electron self-energy of surface state band along $\overline{\Gamma}$--$\overline{M}$ direction.
}
\end{figure}

\newpage

\begin{figure}[tbp]
\begin{center}
\includegraphics[width=1\columnwidth,angle=0]{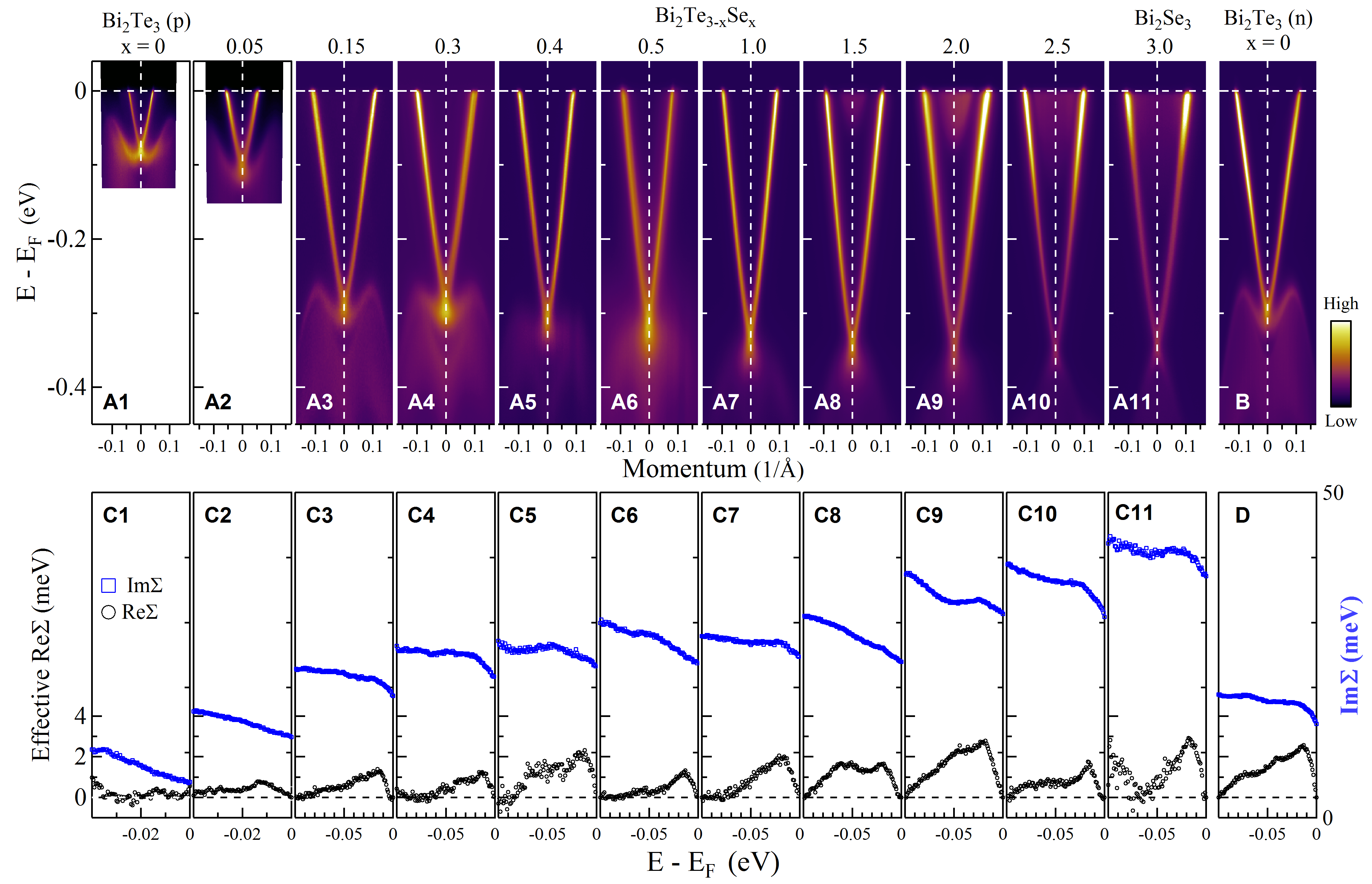}
\end{center}
\caption{Evolution of surface state band structure and electron self-energy with the composition, x, in Bi$_2$Te$_{3-x}$Se$_{x}$ system. (\textbf{A}) The surface state band structure of Bi$_{2}$Te$_{3-x}$Se$_{x}$ with various compositions, x, measured along the $\overline{\Gamma}$--$\overline{K}$ direction. (\textbf{B}) The surface state band structure of n-type Bi$_{2}$Te$_{3}$ along the $\overline{\Gamma}$--$\overline{K}$ direction. (\textbf{C}) Corresponding effective real part of the electron self energy (black circles) and imaginary part of the electron self-energy (blue squares) for Bi$_2$Te$_{3-x}$Se$_{x}$ with different compositions. (\textbf{D}) Corresponding effective real part of the electron self energy (black circles) and imaginary part of the electron self-energy (blue squares) for n-type Bi$_2$Te$_3$.
}
\end{figure}

\newpage

\begin{figure}[tbp]
\begin{center}
\includegraphics[width=1\columnwidth,angle=0]{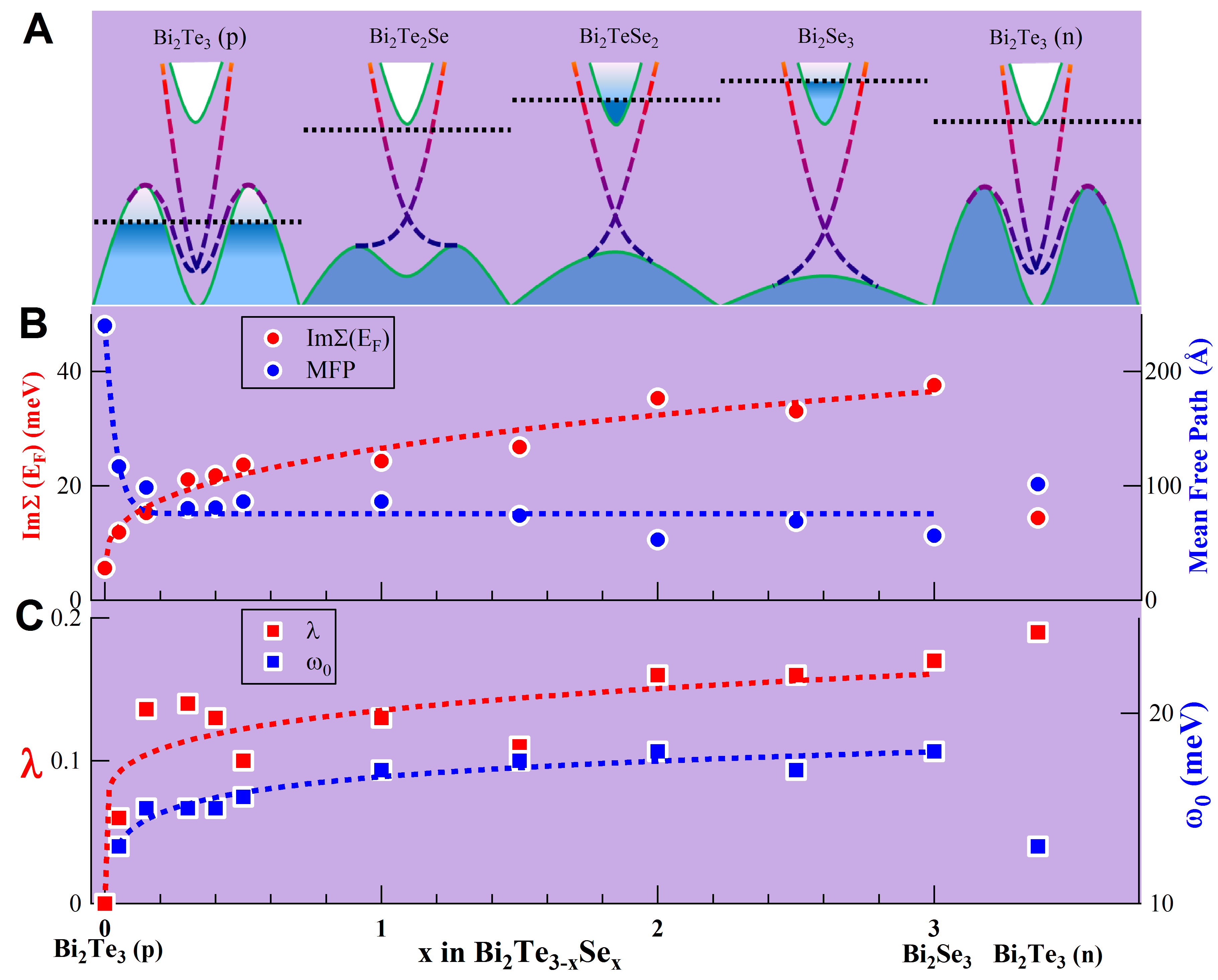}
\end{center}
\caption{Variation of electron-phonon coupling and electron-disorder scattering with composition, x, in  Bi$_{2}$Te$_{3-x}$Se$_{x}$ topological insulators. On the far right side, data on n-type Bi$_2$Te$_3$ are also included. (\textbf{A}) Schematic band structure at several typical compositions to show the relative position between the topological surface states (color dashed lines) and the bulk states (green solid lines). Black dashed lines indicate the Fermi level. (\textbf{B}) Variation of the electron-disorder scattering (red circles) with composition, x, in Bi$_{2}$Te$_{3-x}$Se$_{x}$ topological insulators. It is obtained by taking the value of the imaginary part of the electron self energy at zero energy (E$_F$) as shown in Fig. 3C. Also shown is the electron mean free path (blue circles) calculated from the measured MDC width at the Fermi level.  (\textbf{C}) Variation of the electron-phonon coupling strength $\lambda$ (red squares) and characteristic energy scale $\omega_0$ with the composition, x, in Bi$_{2}$Te$_{3-x}$Se$_{x}$ topological insulators and in n-type Bi$_2$Te$_3$.  The dashed lines in \textbf{B} and \textbf{C} are guide to the eyes.}
\end{figure}

\end{document}